# Thermal conductivity of graphene flakes: Comparison with bulk graphite

**D.L. Nika**[×]**, S. Ghosh, E.P. Pokatilov**[×] **and A.A. Balandin**[∗]

*Nano-Device Laboratory, Department of Electrical Engineering, University of California – Riverside, Riverside, California 92521 USA*

*Materials Science and Engineering Program, Bourns College of Engineering, University of California – Riverside, Riverside, California 92521 USA*

*Abstract*

The authors proposed a simple model for the lattice thermal conductivity of graphene in the framework of Klemens approximation. The Gruneisen parameters were introduced separately for the longitudinal and transverse phonon branches through averaging over phonon modes obtained from the first-principles. The calculations show that Umklapp-limited thermal conductivity of graphene grows with the increasing linear dimensions of graphene flakes and can exceed that of the basal planes of bulk graphite when the flake size is on the order of few micrometers. The obtained results are in agreement with experimental data and reflect the two-dimensional nature of phonon transport in graphene.

---

[×] On leave from the Department of Theoretical Physics, Moldova State University, Chisinau, Republic of Moldova
[∗] Corresponding author; E-mail address: balandin@ee.ucr.edu ; web-address: http://ndl.ee.ucr.edu

It was recently discovered experimentally that graphene, i.e., individual sheets of $sp^2$-hybridized carbon bound in two dimensions (2D) [1-2], reveals an extremely high thermal conductivity $K$. The measurements reported by Balandin et al [3-4] were performed with the non-contact optical technique based on Raman spectroscopy, and used independently determined temperature coefficients of the graphene $G$ phonon peak [5]. It was found that the near room-temperature (RT) thermal conductivity of partially suspended single-layer graphene is in the range $K \sim 3000 - 5000$ W/mK depending on the graphene flake size. The strong width dependence was also observed [4, 6].

The first experimental report of the thermal conductivity of graphene stimulated a body of theoretical work on the subject. Nika et al [7] performed detail study of the lattice thermal conductivity of graphene using the phonon dispersion obtained from the valence-force field (VFF) method. The authors treated the three-phonon Umklapp scattering directly considering all phonon relaxation channels allowed by the energy and momentum conservation in graphene 2D Brillouin zone (BZ) [7]. Jiang et al [8] calculated the thermal conductance of graphene in the pure ballistic limit obtaining a high value, which translates to the thermal conductivity in excess of $\sim 6600$ W/mK. The higher thermal conductivity is expected for the ballistic regime when no scattering is included. Although numerical calculations are in line with the experiments [3-4], there is a strong need for a simple analytical model, which would elucidate the differences in the heat transport in graphene as compared to bulk graphite. Such a model can also be used for estimates of the thermal conductivity of graphene flake of various sizes and at different temperatures, and help with interpretation of the experimental data.

In this letter, we follow the spirit of Klemens' approach to the thermal conductivity of graphene [9-10], which was proposed before graphene was actually exfoliated. We alter it by using a more general expression for thermal conductivity, introducing two Gruneisen parameters $\gamma_s$ obtained independently for each of the heat conducting phonon polarization branches $s$, and keeping separate the velocities and cut-off frequencies for each phonon branch. These allow us to better reflect the specifics of the phonon dispersion in graphene. The effective parameters $\gamma_s$ are computed by averaging the phonon mode-dependent $\gamma(q)$ for all relevant phonons (here $q$ is the phonon wave vector). The phonon branches, which carry heat, are longitudinal acoustic (*LA*) and transverse acoustic (*TA*). The out-of-plane transverse acoustic phonons (*ZA*), i.e. "bending branch", do not make substantial contributions to heat conduction due to their low group velocity and high $\gamma(q)$.

Klemens clearly distinguished the heat transport in basal planes of bulk graphite and in single layer graphene [9-10]. In the former the heat transport is approximately 2D only till some low-bound cut-off frequency $\omega_C$. Below $\omega_C$ there appears strong coupling with the cross-plane phonon modes and heat starts to propagate in all directions, which reduces the contributions of these low-energy modes to heat transport along basal planes to negligible. In bulk graphite there is a physically reasonable reference point for the on-set of the cross-plane coupling, which is the *ZO'* phonon branch near ~4 THz observed in the spectrum of bulk graphite. The presence of *ZO'* branch and corresponding $\omega_C$ allows one to avoid the logarithmic divergence in the Umklapp-limited thermal conductivity integral and calculate it without considering other scattering mechanisms.

The physics of heat conduction is principally different in graphene where the phonon transport is pure 2D all the way to zero phonon frequency $\omega(q=0)=0$. There is no *ZO'* branch in graphene and no on-set of the cross-plane heat transport at the long-wavelength limit in the system, which consists of only one atomic plane. Thus, the cut-off frequency for Umklapp processes can not be introduced by analogy with bulk graphite. In the case of graphene one needs to include other scattering mechanisms for phonons in order to obtain the thermal conductivity. We have previously accomplished it by including phonon scattering on rough edges of graphene flakes and mass-difference defect scattering [7]. One can also avoid divergence by limiting the phonon mean free path (MFP) by the natural boundaries of the crystalline graphene flake.

Using an expression for the three-phonon Umklapp scattering from Refs [9-10] but introducing separate life-times for *LA* and *TA* phonons, we have

$$\tau_{U,s} = \frac{1}{\gamma_s^2} \frac{M v_s^2}{k_B T} \frac{\omega_{s,\max}}{\omega^2}, \qquad (1)$$

where $s=TA, LA$, $v_s$ is the average phonon velocity for a given branch, $T$ is the absolute temperature, $k_B$ is the Boltzmann constant, $\omega_{s,max}$ is the maximum cut-off frequency for a given branch and *M* is the mass of an atom. To determine $\gamma_s$ we averaged $\gamma(q)$ obtained from the accurate phonon dispersion calculated using VFF method [7] and Mounet and Marzari [11] *ab initio* theory. Both approaches give similar results for all phonon branches.

The general expression for the thermal conductivity of graphene can be written as [7]

$$K = \frac{1}{4\pi k_B T^2 h} \sum_{s=TA,LA} \int_{q_{min}}^{q_{max}} \{(\hbar\omega_s(q)\frac{d\omega_s(q)}{dq})^2 \tau_{U,s} \frac{exp[\hbar\omega_s(q)/kT]}{[exp[\hbar\omega_s(q)/kT]-1]^2} q\} dq. \quad (2)$$

Here $h=0.35$ nm is the thickness of graphene and $\tau_{U,s}$ are given by Eq. (1). The above equation can be used to calculate the thermal conductivity with the actual dependence of the phonon frequency $\omega_s(q)$ and the phonon velocity $d\omega_s(q)/dq$ on the phonon wave number. To simplify the model we can use the liner dispersion $\omega_s(q) = \upsilon_s q$ and re-write it as

$$K = \frac{M}{4\pi T h} \sum_{s=TA,LA} \frac{\omega_{s,max} \upsilon_s^2}{\gamma_s^2} F(\omega_{s,min}, \omega_{s,max}), \quad (3)$$

where

$$F(\omega_{s,min}, \omega_{s,max}) = \int_{\hbar\omega_{s,min}/k_BT}^{\hbar\omega_{s,max}/k_BT} x \frac{exp(x)}{[exp(x)-1]^2} dx = [ln\{exp(x)-1\} + \frac{x}{1-exp(x)} - x]\Big|_{\hbar\omega_{s,min}/k_BT}^{\hbar\omega_{s,max}/k_BT}.$$

In the above equation, $x = \hbar\omega/k_B T$, and the upper cut-off frequencies $\omega_{s,max}$ are defined from the actual phonon dispersion [7,11]. The low-bound cut-off frequencies $\omega_{s,min}$ for each *s* are determined from the condition that the phonon MFP cannot exceed the physical size *L* of the flake, i.e.

$$\omega_{s,min} = \frac{\upsilon_s}{\gamma_s} \sqrt{\frac{M\upsilon_s}{k_B T} \frac{\omega_{s,max}}{L}}. \quad (4)$$

The integrand in Eq. (3) can be further simplified near RT when $\hbar\omega_{s,max} > k_B T$, and it can be expressed as

$$F(\omega_{s,\min}) \approx -ln\{|exp(\hbar\omega_{s,\min}/k_B T)-1|\} + \frac{\hbar\omega_{s,\min}}{k_B T} \frac{exp(\hbar\omega_{s,\min}/k_B T)}{exp(\hbar\omega_{s,\min}/k_B T)-1}. \qquad (5)$$

The obtained Eqs. (3) and (5) constitute a simple analytical model for calculation of the thermal conductivity of graphene layers, which retains such important features of graphene phonon spectra as different $\upsilon_s$ and $\gamma_s$ for *LA* and *TA* branches. The model also reflects strictly 2D nature of heat transport in graphene all the way down to zero phonon frequency. Our Eq. (3) reduces to Klemens' formula for graphene [9-10] in the limit $x \to 0$ ($\hbar\omega << k_B T$) and additional simplifying assumption of the same $\gamma_s$ and $\upsilon_s$ for *LA* and *TA* phonons.

Using Eqs. (3-5) we calculated the Umklapp-limited thermal conductivity of graphene as a function of temperature near RT. The results are shown in Fig. 1 for different linear dimensions of graphene flakes. The Gruneisen parameters used in this calculation, $\gamma_{LA}$=1.8 and $\gamma_{TA}$=0.75, were obtained by averaging of $\gamma(q)$ [11]. The calculated Umklapp-limited thermal conductivity of graphene exceeds the experimental RT value for the high-quality bulk graphite, $K\approx$2000 W/mK [12]. It grows with the increasing linear size of the graphene flake *L*. Such dependence may appear surprising near RT. At the same time, it is manifestation of the strictly 2D nature of phonon transport in graphene.

The *K* increase with increasing *L* stems from $\omega_{s,\min} \sim L^{-1/2}$ dependence (see Eq. (4)). This means that in the larger graphene flakes, acoustic phonons with longer wavelength are available for heat transfer. Graphene's low-bound frequencies $\omega_{s,\min}$ fall below the graphite bulk cut-off of 4 THz when the flake size is between 4 – 10 $\mu m$ (depending on temperature). At this length scale, the Umklapp-limited thermal conductivity of graphene exceeds the theoretical intrinsic thermal conductivity of bulk graphite owing to a larger interval of phonons ($\omega_{s,\min}$; $\omega_{s,\max}$) participating in heat conduction (see Eq. (3)). The temperature dependence deviates from *1/T* law, particularly for small flakes, due to the boundary restrictions on phonon MFP from graphene edges. This has analogies in heat transport in polycrystalline materials with similar restrictions on phonon MFP due to the boundaries [13]. The experimental data after Balandin *et al* [3-4] is also shown for comparison. The added error bars indicate the spread of measured values for different sizes of graphene flakes produced from crystalline Kish graphite [3-4, 14].

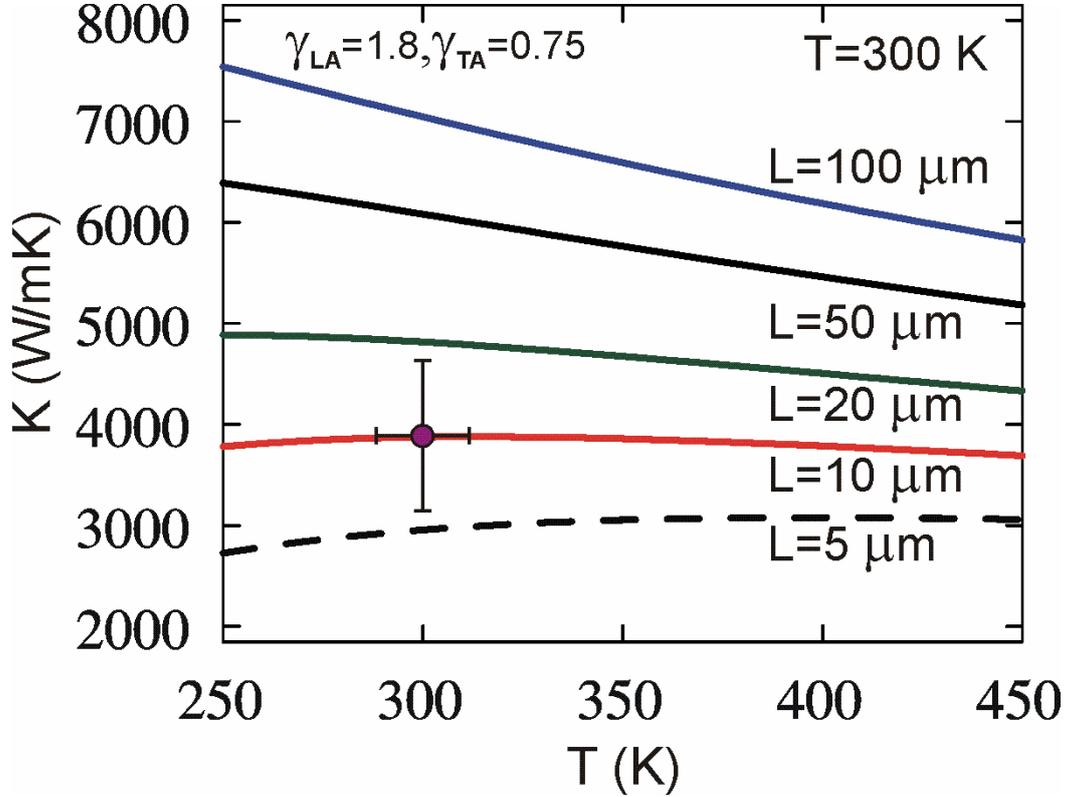

**Figure 1:** Thermal conductivity of graphene flake as a function of temperature for several linear dimensions *L* of the flake.

In Fig. 2 we present the dependence of thermal conductivity of graphene on the dimension of the flake *L*. The data is presented for the averaged values $\gamma_{LA}$=1.8 and $\gamma_{TA}$=0.75 obtained from *ab initio* calculations, and for several other close sets of $\gamma_{LA,TA}$ to illustrate the sensitivity of the result to Gruneisen parameters. For small graphene flakes, *K* dependence on *L* is rather strong. It weakens for flakes with *L*≥10 μm. The calculated values are in agreement with the experiment [3-4, 6]. The horizontal line indicate the experimental thermal conductivity for bulk graphite [12], which is lower than the theoretical intrinsic limit, and it is exceeded by graphene's thermal conductivity at smaller *L*. Klemens' formula [9-10] gives similar *K(L)* dependence but with different absolute values of *K* due to overestimated $\gamma$ in his calculations and few other simplifying assumptions.

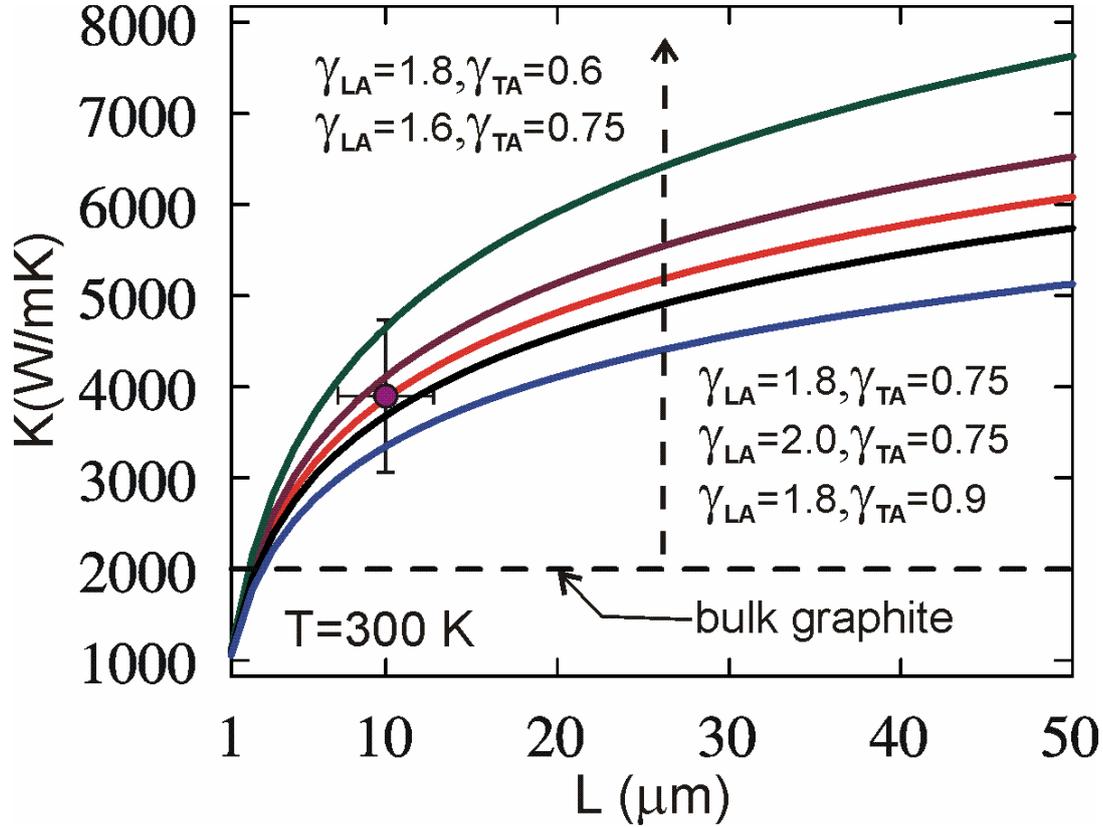

**Figure 2:** Thermal conductivity of graphene as a function of the graphene flake size *L*. Note that the thermal conductivity of graphene exceeds that of basal planes of graphite when the flake size is larger than few micrometers.

One should note here that the calculated thermal conductivity is an *intrinsic* quantity limited by the three-phonon Umklapp scattering only. But it is determined for a specific graphene flake size since *L* defines the low-bound (long-wavelength) cut-off frequency in Umklapp scattering through Eq. (4). In experiments, thermal conductivity will also be limited by defect scattering. When the size of the flake becomes very large with many polycrystalline grains, the scattering on their boundaries will also lead to phonon relaxation. The latter can be included in our model through adjustment of *L*. The extrinsic phonon scattering mechanisms prevent indefinite growth of thermal conductivity of graphene with *L*.

In conclusion, we proposed a simple analytical model for estimating the thermal conductivity of graphene flakes of different size. It captures the main features of pure 2D phonon transport in graphene, which distinguishes it from that in basal planes of bulk graphite. The model utilizes two Gruneisen parameters $\gamma_s$ for the heat conducting *LA* and *TA* phonon branches, which were obtained by averaging of the *ab initio* mode-dependent $\gamma(q)$. The thermal conductivity calculated with our model gives results consistent with the rigorous theory of heat conduction in graphene [7] and in excellent agreement with experiment [3-4]. Obtained results are important for the proposed graphene applications as lateral heat spreaders [4, 6, 14-15] and interconnects [16] in future nanoelectronic and optoelectronic circuits.

*Acknowledgements*

The work was supported, in part, by DARPA – SRC through the FCRP Center on Functional Engineered Nano Architectonics (FENA) and Interconnect Focus Center (IFC).